\documentclass[prd, twocolumn, floatfix]{revtex4}
\usepackage{amsmath}
\usepackage{graphicx}
\usepackage{dcolumn}
\usepackage{bm}
\usepackage{epsfig}
\usepackage{amssymb,latexsym,mathrsfs}
\usepackage{graphicx}
\usepackage{color}
\usepackage{hyperref}

\hypersetup{
    colorlinks=true,
    linkcolor=red,
    citecolor=blue,
}

\newcommand{\be}{\begin{equation}}
\newcommand{\ee}{\end{equation}}
\newcommand{\bea}{\begin{eqnarray}}
\newcommand{\eea}{\end{eqnarray}}

\newcommand{\om}{\Omega_m}

\newcommand{\sigf}{\sigma_f}
\newcommand{\lcdm}{$\Lambda$CDM}

\begin{document}
\title{Gaussian Process Cosmography} 
\author{Arman Shafieloo$^1$, Alex G.\ Kim$^2$, Eric V.\ Linder$^{1,2,3}$} 
\affiliation{$^1$ Institute for the Early Universe WCU, Ewha Womans 
University, Seoul, Korea} 
\affiliation{$^2$ Lawrence Berkeley National Laboratory, Berkeley, CA 94720, 
USA} 
\affiliation{$^3$ University of California, Berkeley, CA 94720, USA}

\begin{abstract}
Gaussian processes provide a method for 
extracting cosmological information from observations without assuming a cosmological model.  We carry 
out cosmography -- mapping the time evolution of the cosmic expansion -- 
in a model-independent manner using kinematic variables and a geometric 
probe of cosmology.  Using the state of the art supernova distance data 
from the Union2.1 compilation, we 
constrain, without any assumptions about dark energy parametrization or 
matter density, the Hubble parameter and deceleration parameter as a 
function of redshift.  Extraction of these relations is tested successfully 
against models with features on various coherence scales, subject to 
certain statistical cautions. 
\end{abstract}

\date{\today} 

\maketitle

\section{Introduction} 

Cosmic acceleration is a fundamental mystery of great interest and 
importance to understanding cosmology, gravitation, and high energy 
physics.  The cosmic expansion rate is slowed down by gravitationally 
attractive matter and sped up by some other, unknown contribution to 
the dynamical equations.  While great effort is being put into identifying 
the source of this extra dark energy contribution, the overall expansion 
behavior also holds important clues to origin, evolution, 
and present state of our universe. 

Indeed, by studying the expansion as a whole one sidesteps the issue 
of exactly how to divide the gravitationally attractive (e.g.\ the 
imperfectly known matter density) and the accelerating contributions, 
and whether 
they are independent or have some interaction (see, e.g., 
\cite{kunz,kunzlid,cervantes}).  By concentrating on the 
kinematic variables -- the expansion properties as a function of redshift 
$z$ or scale factor $a=1/(1+z)$ -- one does not need to know the internal 
structure of the field equations, i.e.\ the dynamics.  The clarity and 
focus on kinematics trades off against the loss of information on the 
specific dynamics. 

Another gain comes from using geometric measurements -- cosmic-distance 
variables that do not depend on the particular forces and mass densities. 
The sensitivity of Type Ia supernova measurements of the distance-redshift 
relation to the deceleration parameter was used to discover the accelerated 
expansion \cite{perl99,riess98}.  Other probes such as gravitational 
lensing, galaxy-clustering 
statistics, cluster-mass abundances, etc.\ provide valuable information, 
but are dependent on non-kinematic variables.  Some techniques, such as 
distances from baryon acoustic oscillations and Sunyaev-Zel'dovich 
effects in clusters, are on the fence, nominally geometric but having 
implicit dependence on the gravitational interaction of matter and so the 
force law dynamics. 

Given distance and redshift measurements, the cosmic expansion rate is 
related by a derivative of the data, and the deceleration parameter by 
a further derivative.  This is problematic for data with real world noise, 
as the differentiation further amplifies the noise.  Various smoothing 
procedures have been suggested, e.g.\ \cite{smoothings}, but tend to 
induce bias 
in the function reconstruction due to parametric restriction of the 
behavior or to have poor error control.  Using a general orthonormal basis 
or principal component analysis is another approach, to describe the 
distance-redshift relation (e.g.\ \cite{11090873}) or 
the deceleration parameter \cite{shatur}, or using a correlated prior for 
smoothness on the dark energy equation of state \cite{crittenden}, but in 
practice a finite (and small) number of modes is significant beyond the 
prior, essentially reducing to a parametric approach.  Gaussian processes 
\cite{gpml} offer an interesting possibility for improving this situation.  

Gaussian processes (GP) have been used recently \cite{la1,la2,la3} in a 
dynamical reconstruction, going from a set of realizations of the equation 
of state parameter $w(z)$ of the dark energy component forward to 
comparison of the derived distance-relation to the distance data.  
The comparison was carried out through a Markov Chain Monte Carlo (MCMC) 
assessment of likelihoods.  Note that the GP interpolation does not occur 
between data points but rather on an arbitrary grid of some possibly 
unmeasured quantity.  This approach is intriguing, but relies on separation 
of the matter density from the dark energy behavior, i.e.\ it works within 
a dynamical framework. 

The approach in this paper takes a fundamentally different path.  We 
begin with the observations of supernova distances and here consider 
only kinematic quantities.  Modeling the cosmic distance relation as a 
smooth kinematic function drawn from a GP, the value of the function at any 
redshift is then predicted directly through testing the GP model against 
the data.  The cosmic expansion can then be extracted from the means 
and covariance matrices of the Gaussian process realizations (weighted by a posterior) directly 
for quantities related linearly to the original GP, even through derivatives.  
This allows us to probe the rate and acceleration of the cosmic 
expansion in a highly model-independent manner (at the price of focusing 
on only this type of information).  One can view this as a top-down 
approach, complementary with the bottom-up approach of starting with 
theoretical quantities and working toward the data, and then applying a 
likelihood comparison.  

In Sec.~\ref{sec:methods} we lay out the basics of the kinematic cosmology 
quantities and the Gaussian process formalism.  Readers familiar with the 
method or eager for results could go to Sec.~\ref{sec:results} where we 
analyze the results of performing the GP reconstruction, for both 
current and simulated data.  We summarize the cosmological implications 
and discuss the prospects in Sec.~\ref{sec:concl}.  Appendices present 
details of tests of the robustness of the statistical techniques.

\section{Cosmographic Reconstruction} \label{sec:methods}

\subsection{Expansion History} \label{sec:exp}

Homogeneity and isotropy determine the metric of the universe to be of 
the Robertson-Walker form, which for a spatially flat universe (from a 
theory or inflationary prior) is 
\be 
ds^2=-dt^2+a^2(t)\, 
\left[dr^2+r^2(d\theta^2+\sin^2\theta\,d\phi^2)\right]\,,  
\ee 
where $t$ is a time coordinate and $r$ the coordinate distance.  The 
key quantity is the cosmic expansion or scale factor $a(t)$, or 
equivalently the redshift $z=a^{-1}-1$. 

Without using any field equations, such as the Friedmann equations (the 
Einstein equations specialized to the above metric) -- and hence in a 
purely kinematic way -- we can still define a conformal distance 
\be 
\eta\equiv \int \frac{dt}{a}=\int dr \,, 
\ee 
and build a luminosity distance $d_L(a)=a^{-1}\eta(a)$, and an angular 
diameter distance $d_a(a)=a\eta$ if desired. 

Flux and redshift measurements of a set of standardized candles such as 
Type Ia supernovae deliver observational access to $\eta(z)$. 
From this 
function directly comes the inverse Hubble parameter 
\be 
H^{-1}(z)\equiv \left(\frac{\dot a}{a}\right)^{-1}=\frac{d\eta}{dz} 
\label{hinv:eqn}
\ee  
and the deceleration parameter 
\be 
q(z)=-\frac{a\ddot a}{\dot a^2}=-\frac{1+z}{H^{-1}}\frac{dH^{-1}}{dz}-1 \ . 
\label{eq:qdef} 
\ee 
This a top-down approach, starting with observable distances and proceeding 
to cosmological kinematic quantities. 

We follow this top-down approach because of its useful properties of 
direct relation to kinematics, avoidance of reliance on a cosmological 
model or knowledge of the matter density, and well defined and efficient 
error propagation within the GP method. 

An alternative approach with different characteristics is bottom-up.  
There, one would either parametrize $q(a)$ (or 
perhaps a dark energy equation of state, or pressure to density, 
ratio $w_{de}(a)=[2q(a)-1]/[3\Omega_{de}(a)]$, which involves the 
dimensionless dark energy density $\Omega_{de}$) or choose 
realizations of $q(a)$ from a statistical distribution.  Parametrizing 
$q(a)$ allows straightforward error propagation up to the distances, 
for comparison to the data; however one must ensure that the 
parametrization does not restrict or bias the results.  Note that 
choosing a form $q(a)$ is an explicitly dynamical assumption, breaking 
the kinematic nature of the analysis \cite{linrpp}.  In terms of the equation 
of state, the $w_0$-$w_a$ form $w_{de}(a)=w_0+w_a(1-a)$ is highly robust, 
reconstructing 
$d_L(a)$ to better than 0.1\% for a wide array of models \cite{calde} 
but does require a separation into matter density and dark energy behavior. 

If one uses statistical realizations of $q(z)$, then the error 
propagation necessary, including the covariances between values $i$ at 
different redshifts, is 
\be  
\begin{split} 
\mbox{Cov}[H_i,H_j]&=H_i H_j \times\\ 
&\ \int_0^{z_i}\frac{dz'}{1+z'}\,\int_0^{z_j}
\frac{dz''}{1+z''}\,\mbox{Cov}[q_i,q_j]\\ 
\mbox{Cov}[d_i,d_j]&=(1+z_i)(1+z_j) \times \\
&\ \int_0^{z_i}\frac{dz'}{H(z')}
\int_0^{z_j}\frac{dz''}{H(z'')}\,\mbox{Cov}[H_i,H_j] \,. 
\end{split} 
\ee  
This can be slow numerically, especially in a MCMC likelihood evaluation. 

However, for a GP the relation between the covariance of a quantity and 
its derivative (as we use in the top-down approach) is particularly simple 
and furthermore one can avoid functional parametrizations or statistical 
distributions of the cosmological variables.

\subsection{Gaussian Process Modeling} 
\label{sec:gpmethod} 

We begin with the assumption that the stochastic data is described 
by a Gaussian process that corresponds to the cosmological function 
$\eta(z)$.
The effective supernova magnitudes at peak brightness, $m$, and their associated covariance are derived from light-curve data 
\citep[e.g.][]{2007A&A...466...11G}.
Those peak  magnitudes  transformed by $(1+z)10^{m/5}$ represent measurements of the conformal distance with
a nuisance normalization factor,
$y(z)=10^{M/5} \left(10 \mbox{pc}\right)^{-1}\eta(z)$ where $M$ is 
the absolute supernova magnitude.
 
Derivatives of Gaussian processes are themselves Gaussian 
processes (with some ignorable pathological exceptions).  This means 
that the estimator for the Hubble length $H^{-1}(a)$ is also a GP (this 
does not hold for nonflat universes).  
The deceleration parameter is not a GP because of its nonlinear relation 
to $H^{-1}$ but its mean value and covariance can estimated analytically 
from the two GP functions, $dy/dz$ and $d^2y/dz^2$, that it depends on.  
That is, 
\begin{align} 
H^{-1}(a) &\propto \frac{dy}{dz} \\ 
q(a)&=-(1+z)\left(\frac{dy}{dz}\right)^{-1}\frac{d^2y}{dz^2}-1 \ . \label{eq:qy} 
\end{align} 

\subsubsection{Gaussian Process of the Kinematic Function} \label{sec:gpkin} 

The Gaussian processes serve as a regression tool to infer directly from 
distance data the kinematic expansion properties as a function of redshift 
$z$.  This provides the covariances between the values at different 
redshifts as well, which one would expect a physical function to have.  

A Gaussian process is defined as a collection of random variables, any 
finite number of which have a joint Gaussian distribution \cite{gpml}.  
A GP $f(z)$ is specified by a mean function $m(z)$ and a covariance 
function or kernel $k(z,z')$.  For a finite set $Z$ of $z$'s, values of 
the function are drawn from a Normal distribution, 
$\mathbf{f} \sim \mathcal{N}\left(m(Z),K(Z,Z)\right)$ where the matrix 
element $K_{ij}=k(Z_i,Z_j)$. 

The mean function $m(z)$ is an initial guess for the function, 
in effect ``pre-whitening'' the data to reduce the dynamic range over 
which the variations need to be fit.  Without sufficient care the results 
can in fact be influenced by the mean function, so for example assuming 
a $\Lambda$CDM concordance relation is not necessarily a good choice.  We 
discuss the issues in Appendix~\ref{sec:appmean} where we compare several 
approaches to choosing a mean function and investigate their influence.  
In the main text we adopt an iterated smoothings set for the mean function 
and verify that the final results are not influenced by this input. 

For the covariance function we use a common form, the squared 
exponential \footnote{$K$ 
must be positive definite for any set of $z$'s.  
The squared exponential 
form is at the limit of this condition so we actually use an exponent of 
1.999 rather than 2, as did \cite{la1}.}, 
\be 
k(z,z')=\sigf^2\,\exp{\left(-\frac{|z-z'|^2}{2l^2}\right)}, 
\ee 
where $\sigf$ defines the overall amplitude of the correlation (one can 
think of this as an offset or tilt of the reconstructed function from the 
input mean function), and $l$ gives a measure of the coherence length of the 
correlation.  
These effects are discussed and illustrated in 
Appendix~\ref{sec:apphyper}. 

Any parameters for the mean function (such as fiducial $\om$ and $w$, which 
we do not use), and 
$\sigma_f^2$ and $l$, are 
hyperparameters in the fit.

\subsubsection{Gaussian Process of the Data} \label{sec:gpdata} 

In addition to the regression variation represented by the GP covariance 
function, in the data there is intrinsic dispersion in the distance 
indicator and (possibly correlated) measurement noise.  The sum of all 
these gives the GP of the measured data $\mathbf{y}$, with covariance function 
\begin{equation}
k_y(z_i,z_j)=k(z_i,z_j)+\sigma^2_I\,\delta_{ij}+N(\mu_i,\mu_j) \,,
\label{kmu:eqn}
\end{equation} 
where $\sigma^2_I$ is the intrinsic dispersion and $N(\mu_i,\mu_j)$ is 
the measurement noise covariance matrix.

Note that because $q(z)$ involves a ratio of distance derivatives, it is 
immune to the absolute amplitude of the distance, i.e.\ $H_0$ or its 
combination with the absolute supernova magnitude $\mathcal{M}$.  In using 
the smoothing method to generate the mean functions for the GP (see 
Appendix~\ref{sec:appmean}) we fit out the absolute amplitude, making 
the kinematics independent of these nuisance parameters.  
Fitting for $\mathcal{M}$ is a key step that should not be 
neglected, and its uncertainties must be propagated into the final 
reconstruction uncertainties.  Fixing it to a particular value can also 
bias the results.  The smoothing method has been shown robust for including 
$\mathcal{M}$ in \cite{Shafieloo07}, and a similar 
approach has been used in \cite{Crossing3} 
in reconstruction of the expansion history of the universe by combining 
a smoothing method and Crossing Statistic \cite{Crossing1,Crossing2}.  

\subsubsection{Inferring Kinematic Functions from Data} \label{sec:kindata}

Given data $\mathbf{y}$  measured at a set of points $Z$ we want a 
faithful reconstruction of the distance $\eta$, as 
well as its derivatives, at some other set of points $Z_1$.  Call the 
reconstructed function $\mathbf{f}$.  
In GP, the joint probability distribution is given by 
\begin{equation}
\left[
\begin{array}{c}
\mathbf{y} \\
\mathbf{f}
\end{array}
\right]
\sim
\mathcal{N}
\left(
\left[
\begin{array}{c}
\mathbf{m(Z)} \\
\mathbf{m(Z_1)}
\end{array}
\right]
,
\left[
\begin{array}{cc}
K_y (Z,Z) & K (Z,Z_1)\\
K (Z_1,Z)&  K (Z_1,Z_1)
\end{array}
\right]
\right) \,.
\end{equation}
Here the subscript $y$ is just to clearly indicate the GP of the input data. 
The conditional distribution of $\mathbf{f}$ given the data
is described by 
\begin{align}
\overline{\mathbf{f}} & =  m(Z_1)+K(Z_1,Z) K^{-1}_y (Z,Z)\, \mathbf{y} \label{mn:eqn} \\
\mbox{Cov}\left(\mathbf{f}\right) & =K (Z_1,Z_1) - K(Z_1,Z) K^{-1}_y (Z,Z) K(Z,Z_1) \,. \label{cov:eqn} 
\end{align} 

The probability distribution functions (PDFs) of the reconstructed 
functions (see Eq.~\ref{likelihood:eqn} for details) 
are integrated over the hyperparameter space, weighted by the 
hyperparameter posterior distribution.  For each point in hyperparameter 
space the PDF of the GP function and its derivatives (e.g.\ the 
distance, Hubble length, and second derivative entering the deceleration 
parameter) can be written analytically: 
\begin{widetext}
\begin{equation}
\left[
\begin{array}{c}
\mathbf{y} \\
\mathbf{f} \\
\mathbf{f'}\\
\mathbf{f''}
\end{array}
\right]
\sim
\mathcal{N}
\left(
\left[
\begin{array}{c}
\mathbf{m(Z)} \\
\mathbf{m(Z_1)} \\
\mathbf{m'(Z_1)}\\
\mathbf{m''(Z_1)}
\end{array}
\right]
,
\left[
\begin{array}{ccc}
\Sigma_{00}(Z,Z) & \Sigma_{00}(Z,Z_1)  &  \Sigma_{01}(Z,Z_1) \hspace{3mm}   \Sigma_{02}(Z,Z_1) \\
\Sigma_{00}(Z_1,Z) & \Sigma_{00}(Z_1,Z_1)  &  \Sigma_{01}(Z_1,Z_1) \hspace{3mm}  \Sigma_{02}(Z_1,Z_1) \\
\Sigma_{10}(Z_1,Z) & \Sigma_{10}(Z_1,Z_1) &  \Sigma_{11}(Z_1,Z_1) \hspace{3mm}   \Sigma_{12}(Z_1,Z_1)\\
\Sigma_{20}(Z_1,Z)  & \Sigma_{20}(Z_1,Z_1)  &  \Sigma_{21}(Z_1,Z_1)  \hspace{3mm}  \Sigma_{22}(Z_1,Z_1)
\end{array}
\right]
\right) ,
\end{equation} 
\end{widetext}
where 
\begin{equation}
\Sigma_{\alpha\beta} = \frac{d^{\left(\alpha+\beta\right)}K}{dz_i^\alpha dz_j^\beta} \ , 
\end{equation} 
and a prime indicates $d/dz$. 

The inferred mean and covariance of the derivatives are given by
\begin{equation}
\left[
\begin{array}{c}
\overline{\mathbf{f}}\\
\overline{\mathbf{f'}}\\
\overline{\mathbf{f''}}
\end{array}
\right]
=
\left[
\begin{array}{c}
\mathbf{m(Z_1)}\\
\mathbf{m'(Z_1)}\\
\mathbf{m''(Z_1)}
\end{array}
\right]
+
\left[
\begin{array}{c}
\Sigma_{00}(Z_1,Z) \\
\Sigma_{10}(Z_1,Z)\\
\Sigma_{20}(Z_1,Z)
\end{array}
\right]
\Sigma_{00}^{-1}(Z,Z)  \,\mathbf{y}
\end{equation}
\begin{widetext}
\begin{equation}
\mbox{Cov}\left(
\left[
\begin{array}{c}
\mathbf{f}\\
\mathbf{f'}\\
\mathbf{f''}
\end{array}
\right]
\right)
=
\left[
\begin{array}{cc}
\Sigma_{00}(Z_1,Z_1) & \Sigma_{01}(Z_1,Z_1) \hspace{2mm} \Sigma_{02}(Z_1,Z_1)  \\
\Sigma_{10}(Z_1,Z_1) & \Sigma_{11}(Z_1,Z_1) \hspace{2mm} \Sigma_{12}(Z_1,Z_1)  \\
\Sigma_{20}(Z_1,Z_1) & \Sigma_{21}(Z_1,Z_1) \hspace{2mm} \Sigma_{22}(Z_1,Z_1) 
\end{array}
\right]
-
\left[
\begin{array}{c}
\Sigma_{00}(Z_1,Z) \\
\Sigma_{10}(Z_1,Z)\\
\Sigma_{20}(Z_1,Z)
\end{array}
\right]
\Sigma_{00}^{-1}(Z,Z)
\left[\Sigma_{00}(Z,Z_1),\Sigma_{01}(Z,Z_1),\Sigma_{02}(Z,Z_1)\right].
\end{equation}
\end{widetext}
For $q(z)$, which is not a GP, the mean is given by 
Eq.~(\ref{eq:qy}) and its variance is 
\begin{equation}
\begin{split}
\mbox{Var}[q(z)]= (q&+1)^2\left[\frac{\mbox{Var}[y'(z)]}{y'^2(z)} \right.\\ 
&+\left.\frac{\mbox{Var}[y''(z)]}{y''^2(z)}-2\frac{\mbox{Cov}[y'(z),y''(z)]}{y'(z)y''(z)}\right] \ .  
\end{split} 
\end{equation}

To integrate over the hyperparameter space (with its non-Gaussian posterior) 
we can either perform a Monte Carlo integration or do grid sampling.  
Since we only have two hyperparameters, $\sigf^2$ and $l$, we use grid 
sampling, equispaced in logarithm with priors of 
$10^{-5} \leq \sigma_f^2 \leq 1$ and $10^{-2} \leq l \leq 10^{0.2}=1.6$. 
The final 
reconstructed results are weighted averages from the posterior based on 
results from all points in the sampled hyperparameter space.  See 
Appendix~\ref{sec:apphyper} for further details.

\subsection{Data and Simulations} \label{sec:datasim} 

To test the robustness of the reconstruction we perform cosmographic fits 
to simulated data, and then we also perform fits to actual current data.  
For current data we use the Union2.1 supernova compilation, 
including full error covariance matrix, consisting of 580 distances 
from $z=0.02-1.4$.  The simulated data consists of the same number over the 
same range, realizing distances using a random intrinsic dispersion of 6\% 
in distance, for various input cosmologies. 

These different cosmologies are intended both to test the robustness of 
the GP reconstruction and to explore the discriminatory power of the 
reconstruction.  They are summarized in Table~\ref{tab:cos} and all have 
dimensionless present matter density $\om=0.27$.  One is a $\Lambda$CDM 
cosmology.  Another is a member of the family of mirage models \cite{mirage}, 
which match the distance to CMB 
last scattering of $\Lambda$CDM, using the relation for the dark energy 
equation of state $w_a=-3.63(1+w_0)$.  In the limit that $w_0=-1$ this 
family reduces to $\Lambda$CDM.  Note their phantom crossing of 
$w=-1$ can give unusual features in $q(z)$ that are useful for testing the 
GP reconstruction.  The last model is a rapidly evolving dark energy 
cosmology with a sharp transition between a high redshift value of the 
dark energy equation of state parameter $w=-0.5$ and a low redshift 
value $w=-1$.  The specific $w(z)$ is given by the CCL or ``kink'' form 
\cite{cora} with the same parameters used in \cite{la2}.  This is 
a much more rapid transition, and at a lower redshift, than expected 
in general from dark energy, and so also poses a challenging test of 
reconstruction.

\begin{table}[!htb]
\begin{tabular}{l|l} 
Cosmology$\quad$&$\quad$ Description\\ 
\hline 
$\Lambda$CDM&$\quad$ $\om=0.27$, $w=-1$\\
Mirage$\quad$ &$\quad$ $\om=0.27$, $w_0=-0.7$, $w_a=-1.09$\\ 
Kink&$\quad$ $\om=0.27$, $w_0=-1$, $w(z\gg0.5)=-0.5$\\ 
\end{tabular}
\caption{Input cosmologies used to generate simulated distance data 
from which the GP tries to reconstruct the appropriate kinematic 
cosmological quantities.} 
 \label{tab:cos}
\end{table}

\section{Expansion History Results} \label{sec:results} 

From many realizations of the GP reconstruction we reconstruct the 
kinematic functions $\eta$, $H^{-1}=\eta'$, and $q(\eta',\eta'')$, 
where prime denotes $d/dz$, and their PDFs. 
We show error bands at every redshift defined such that 
68\% of the realizations lie within this range.  We emphasize that the 
error band should be interpreted in a redshift by redshift sense and the 
covariances are not visible in such a plot; that is, the upper part of the 
band at one redshift may be correlated with the lower part of the band at 
another redshift. 

\subsection{Simulated Data} 

\begin{figure}[!hbtp] 
  \begin{center}{
\includegraphics[angle=-90,width=\columnwidth]{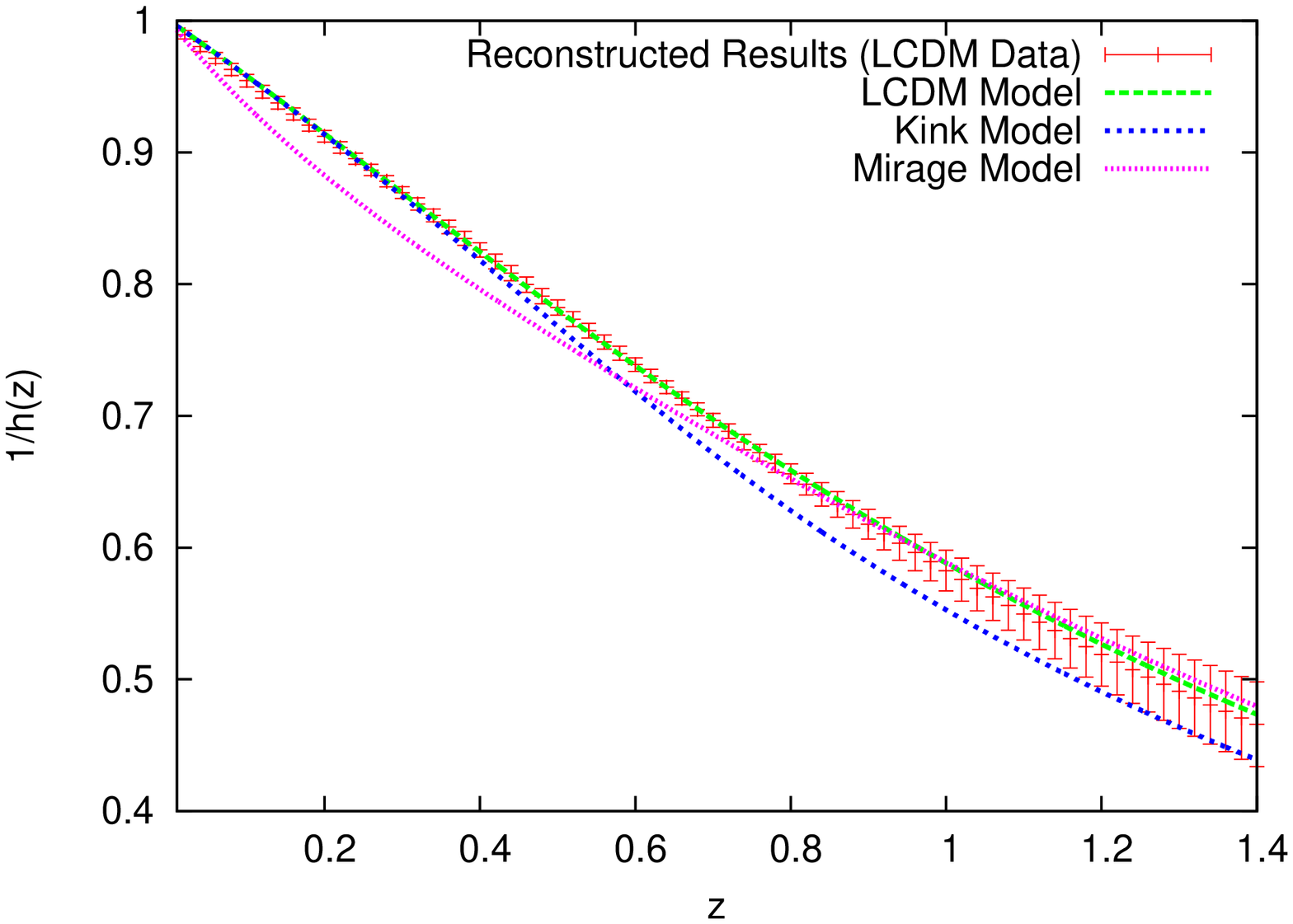}
\includegraphics[angle=-90,width=\columnwidth]{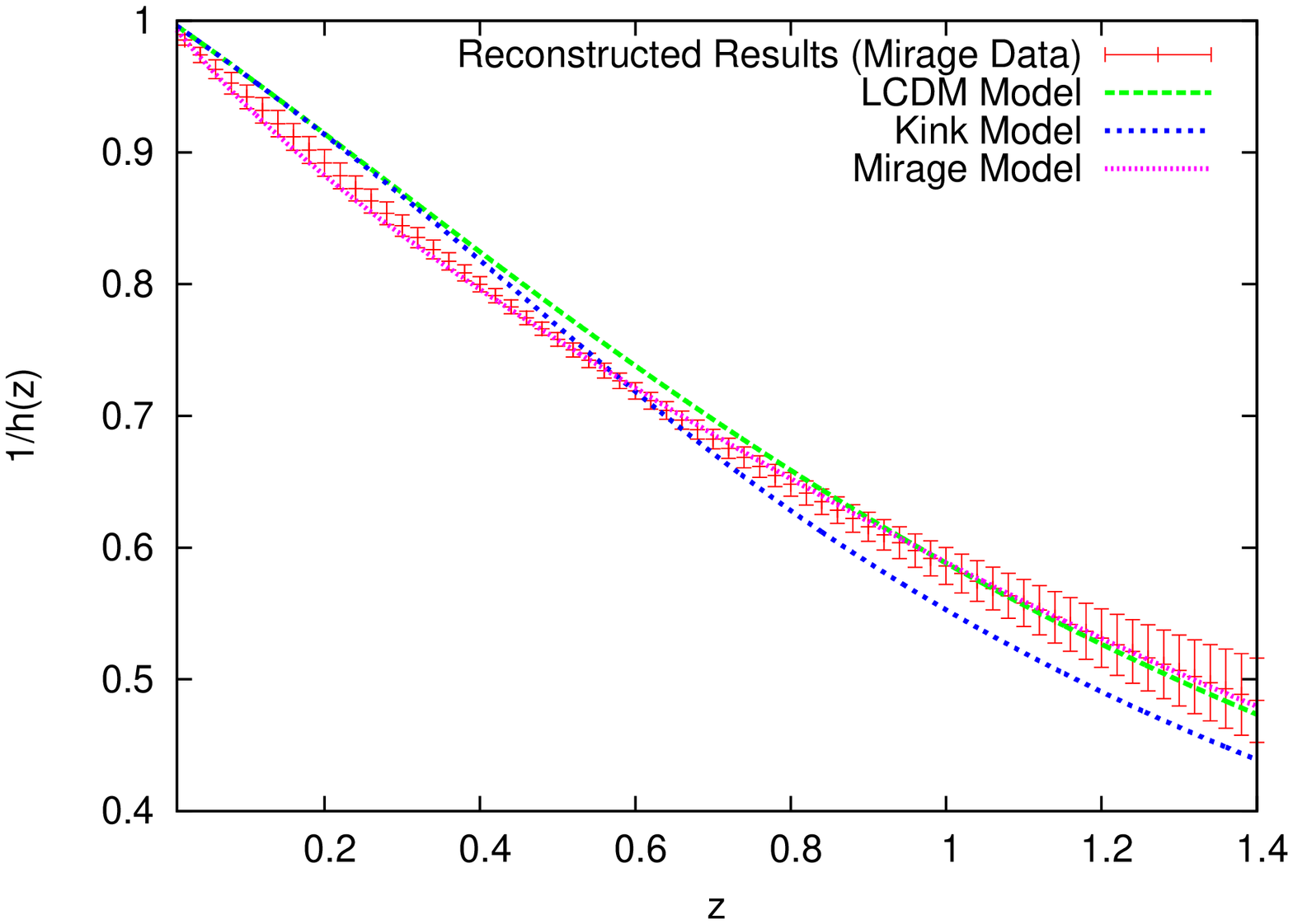}
\includegraphics[angle=-90,width=\columnwidth]{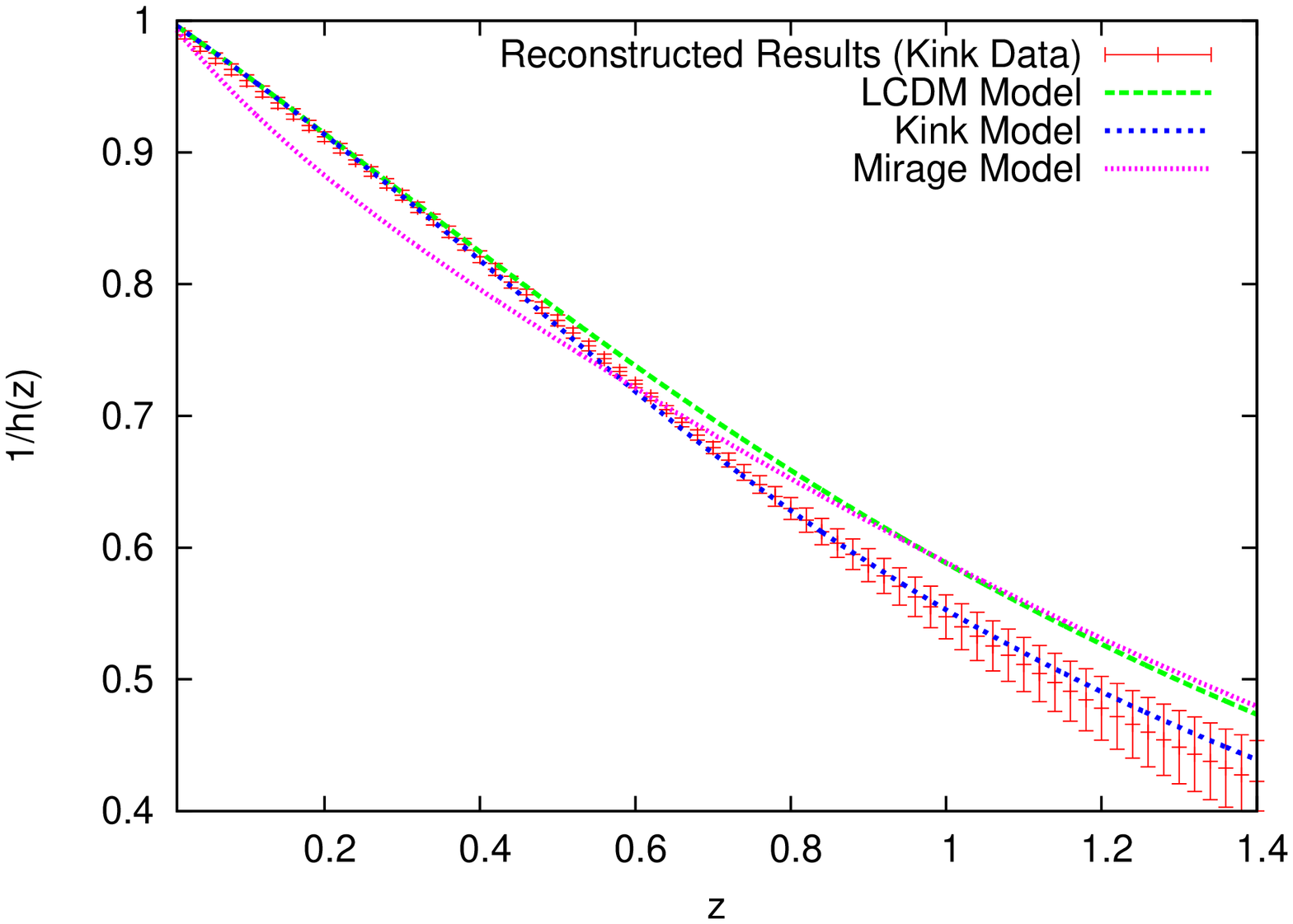}
  }
  \end{center} 
\caption{GP reconstructions of the inverse Hubble parameter $h^{-1}(z)\propto 
H^{-1}(z)$ 
are given for simulated data based on the three different input cosmologies 
of Table~\ref{tab:cos}.  
The dashed curves, the same in all panels, give the true relations.  The 
error band on each reconstruction represents the 68\% confidence level.  
The reconstruction in each case faithfully agrees with the input cosmology. 
}
\label{fig:hrecon}
\end{figure}

\begin{figure}[!hbtp] 
  \begin{center}{
\includegraphics[angle=-90,width=\columnwidth]{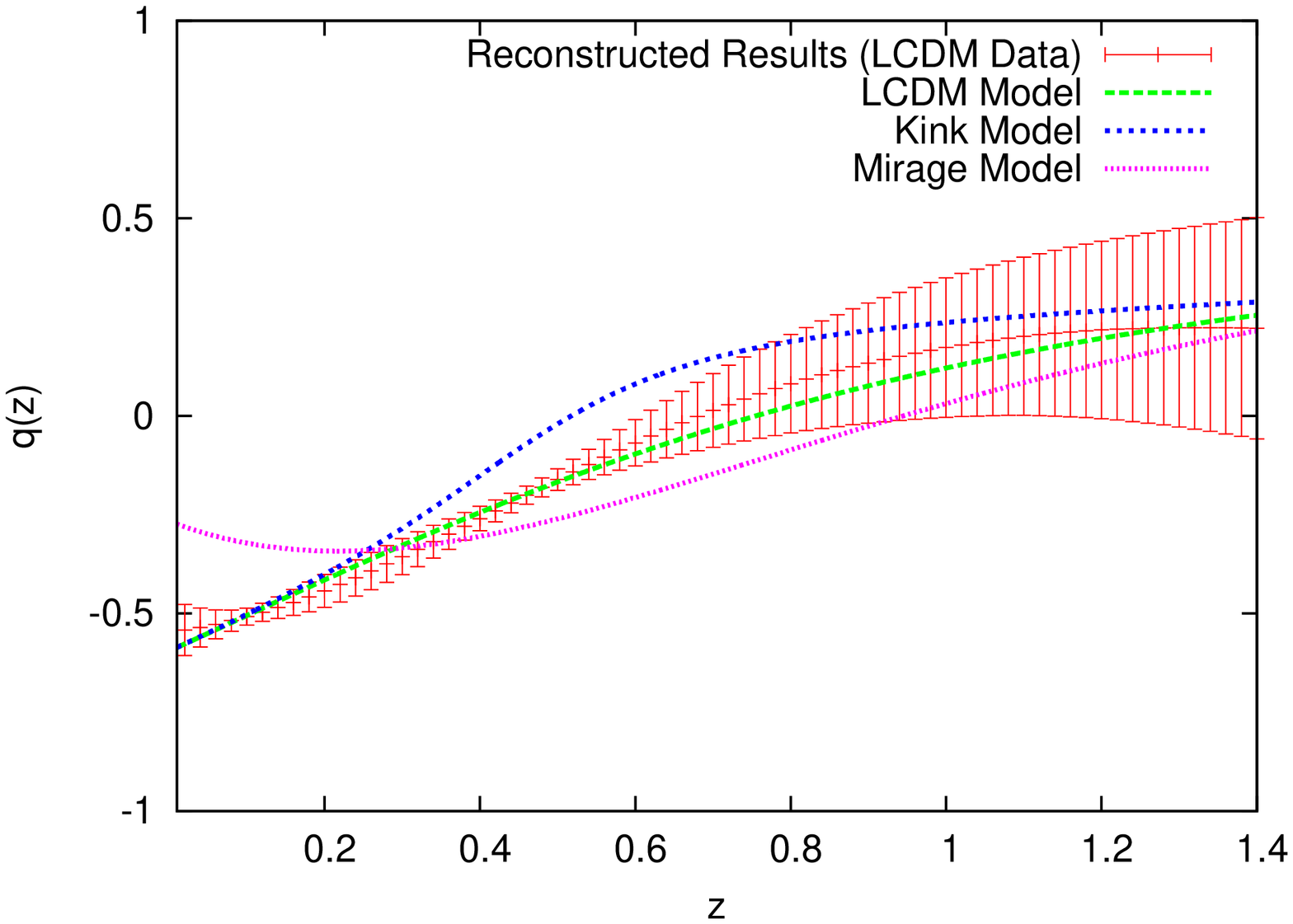}
\includegraphics[angle=-90,width=\columnwidth]{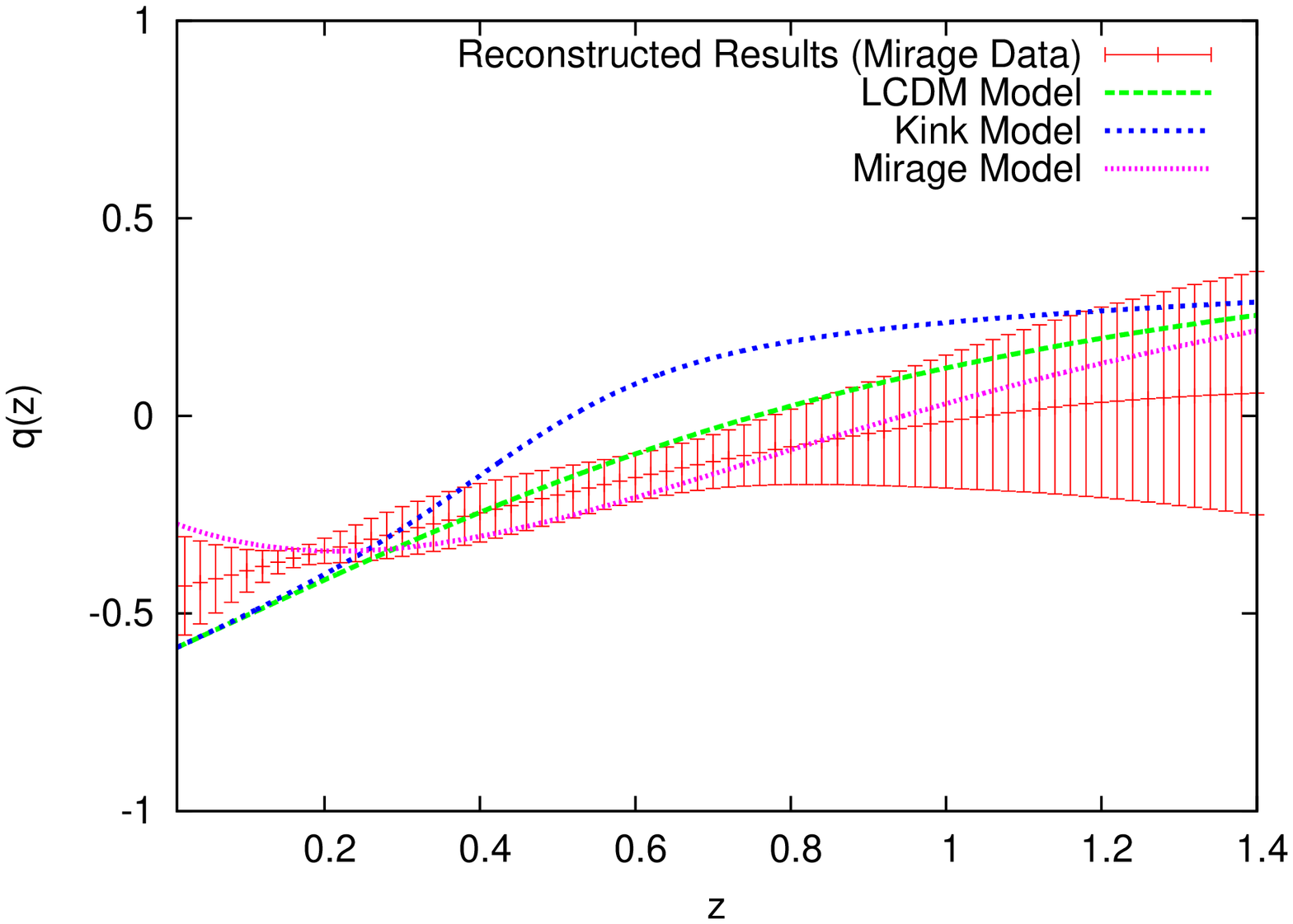}
\includegraphics[angle=-90,width=\columnwidth]{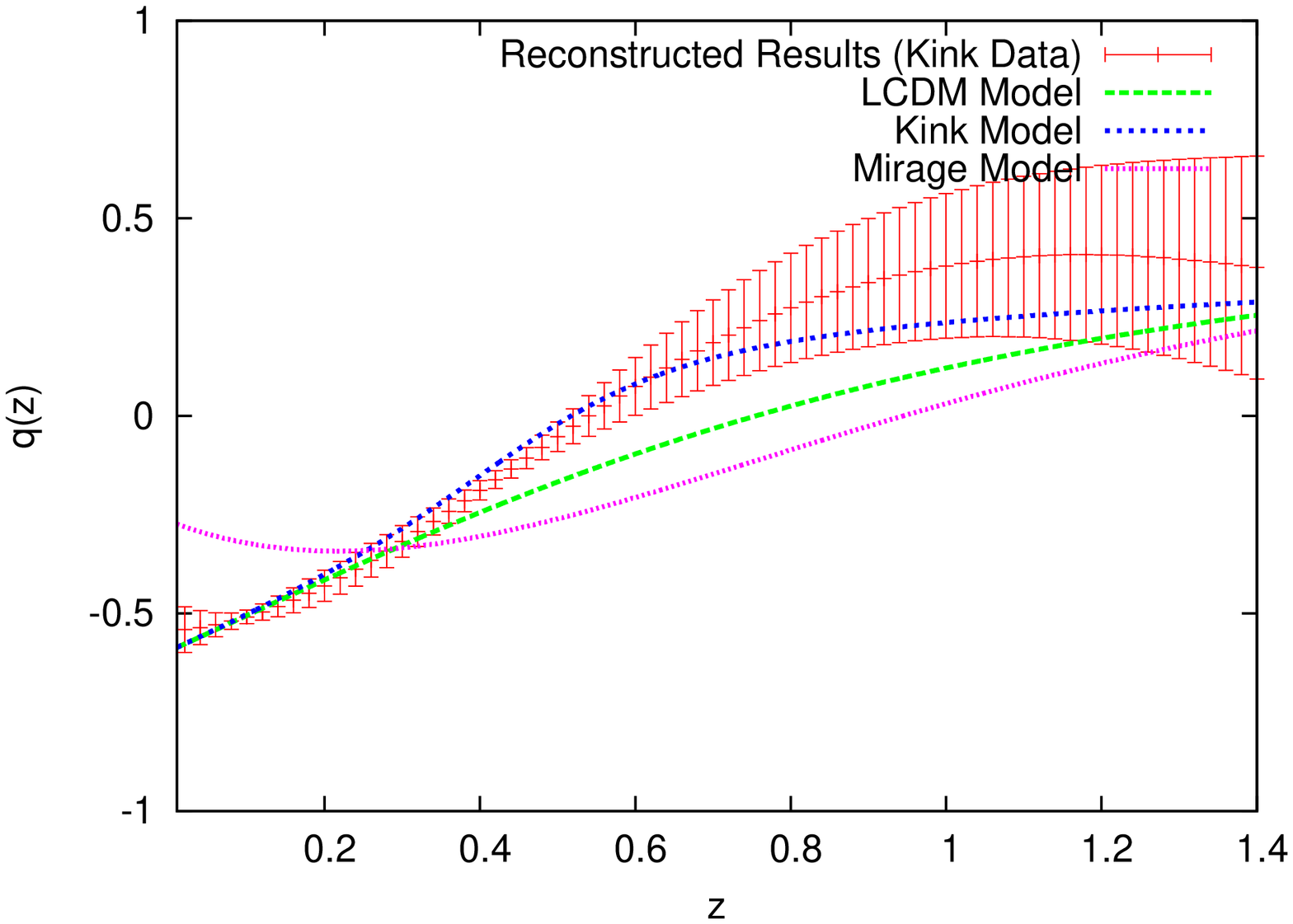}
  }
  \end{center} 
\caption{As Fig.~\ref{fig:hrecon}, but for the deceleration parameter 
$q(z)$. 
}
\label{fig:qrecon}
\end{figure}

For each of the cosmologies in Table~\ref{tab:cos} we plot 
$h^{-1}\propto H^{-1}(z)$ in Fig.~\ref{fig:hrecon} and $q(z)$ 
in Fig.~\ref{fig:qrecon}. 
The true relations are given for each cosmology by the long, short, and 
medium dashed curves (the same in all panels of a set).  The GP 
reconstructions are shown using simulated data based on each cosmology in 
turn, with 68\% error bands.  If the error bands fail to overlap the true 
relation, the GP reconstruction would be inaccurate at 68\% confidence level; 
if the error bands fail to overlap the alternate cosmologies' true 
relations, GP is successful at distinguishing these models.  (Note that 
these are conservative criteria since even if error band overlaps a true 
relation this does not necessarily mean agreement with that cosmology 
because the redshift correlations are not visible.)

We see that each input cosmology is faithfully reconstructed, and each 
alternate cosmology is properly excluded (at 68\% confidence or greater). 
These results demonstrate that GP can be a useful statistical tool for 
model independent kinematic parameter estimation.  

Agreement of the error band with the input model is a necessary but 
not wholly sufficient condition for accurate reconstruction.  One needs 
to take into account the correlations between the predictions at each 
redshift.  Rather than do a model by model, full likelihood computation, 
we tested the influence of correlations 
between redshifts through model independent, simple statistics.  The first 
used the $Om$ function \cite{om} of the Hubble parameter that 
serves as a straightforward consistency test of $\Lambda$CDM, and the second 
example used the deceleration parameter.  Looking at the distribution of 
the differences $\Delta Om(0.2,0.9)\equiv Om(z=0.2)-Om(z=0.9)$ and 
$\Delta q(0.2,0.9)$, for example, we find agreement with the error band 
results that the GP reconstructions accurately reproduce the input 
cosmology values.  Since our analysis assumes no dynamics, we do not have 
to split components into matter and dark energy, and so our results would 
apply to data generated with different input $\Omega_m$ (as we have tested) 
and even cases with coupling between them.

\subsection{Current Data} \label{sec:union21} 

\begin{figure}[!hbtp] 
  \begin{center}{
\includegraphics[angle=-90,width=\columnwidth]{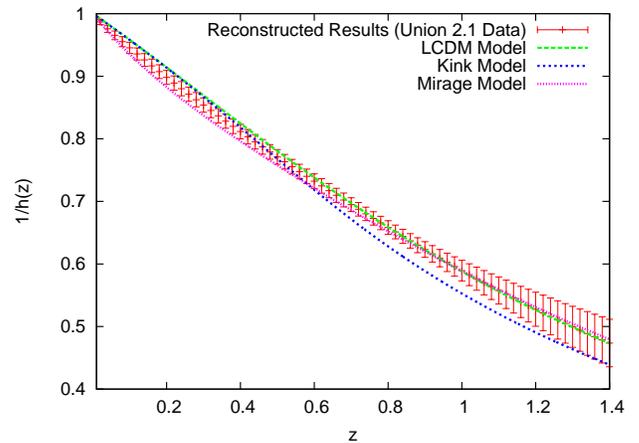}
  }
  \end{center} 
\caption{GP reconstruction of the inverse Hubble parameter $h^{-1}(z)\propto 
H^{-1}(z)$ 
using the Union2.1 data compilation is given by the shaded error band 
representing the 68\% confidence level. 
}
\label{fig:realhrecon}
\end{figure}

\begin{figure}[!hbtp] 
  \begin{center}{
\includegraphics[angle=-90,width=\columnwidth]{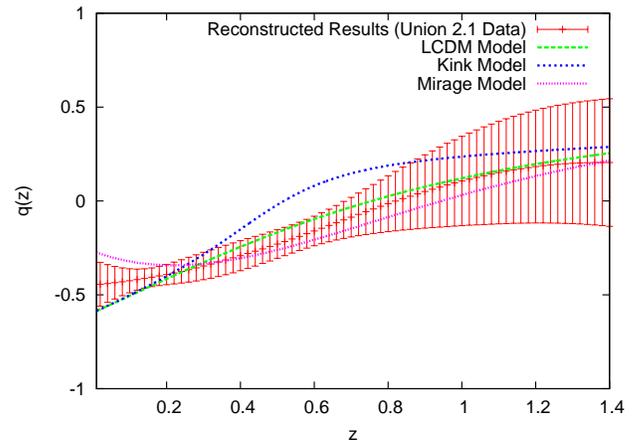}
  }
  \end{center} 
\caption{As Fig.~\ref{fig:realhrecon}, but for the deceleration parameter 
$q(z)$. 
}
\label{fig:realqrecon}
\end{figure}

We now apply the model-independent constraints from GP to the expansion 
history reconstructed from actual current data.  The Union2.1 compilation 
\cite{suzuki} carried out a homogeneous, blind, systematics-oriented 
analysis of supernova distance data.  We use their full data covariance matrix 
for the statistical plus systematics uncertainties.  
The kinematic reconstruction results are presented in 
Figs.~\ref{fig:realhrecon} and \ref{fig:realqrecon}.  

The $\Lambda$CDM model with $\om=0.27$ (now not an input for the mock data, 
but a comparison to 
the fit) is found to be in strong concordance with the Union2.1 data.  

However, we cannot distinguish \lcdm\ from the mirage family of models, 
even one with as extreme time variation as $w_0=-0.7$, $w_a=-1.09$.  Partly 
this is due to the best fit from current data lying between the two, 
roughly corresponding to a mirage model with $w_0=-0.85$, $w_a=-0.54$, and 
partly due to using the full covariance matrix with systematics for current 
data, which gives larger error bars than the simulated statistical errors 
used in the previous plots. 

Current data does point unambiguously to current acceleration in this 
model independent reconstruction, with $q<0$ at low redshift with strong 
significance.  However, current kinematic data does not indicate 
when dark energy fades into the past, i.e.\ $q>0$ is not required at 
$z\gtrsim1$ from this data.

\section{Conclusions} \label{sec:concl}

Gaussian processes can be successfully used in a substantially 
nonparametric reconstruction of the kinematic quantities characterizing 
the cosmic expansion.  We simulated several cosmologies and found that 
GP chose the correct one each time.  GP also has the advantage of simple, 
well controlled propagation of errors and covariances to derivatives (or 
integrals) of the function, allowing distance relations to be converted 
to the Hubble length or second derivative (which can then be analytically 
propagated to the deceleration parameter). 

Key ingredients entering GP are the mean function and covariance 
function, with their hyperparameters.  We emphasized caution in adopting 
such functions that might impose (hidden) restrictions on the reconstruction, 
e.g.\ through the coherence length, possibly leading to results retaining 
memory of the starting point.  Proper treatment of hyperparameters through 
weighted integration over their space and sufficiently wide priors is 
essential.  The combination of smoothing based on the 
data itself, iteration to remove initial conditions, and a set of mean 
functions to enable diversity in GP amplitudes and correlations seems to 
deliver robust results based on our tests.  Future work aims at refining 
this approach further, in particular studying the stability of results 
for a broad range of input cosmologies. 

Our reconstructions of $H^{-1}(z)$ and $q(z)$ using current data are 
consistent with $\Lambda$CDM, but also with cosmologies with substantial 
time variation.  In particular, this holds for the mirage class of models, 
which preserve the distance to CMB last scattering and so the addition of CMB 
data to the supernova data will not affect this conclusion.  Dynamical 
probes, such as growth, in combination with the kinematic distance 
measurements used can have further leverage, but mirage models also have 
substantially similar growth to \lcdm; for example our extreme mirage model 
agrees in growth as a function of redshift to within 1.5\% with \lcdm\ with 
the same matter density.  

While mirage models cross $w=-1$, no conclusions can be drawn from the data 
regarding the necessity for such crossing to occur.  Furthermore, until 
future data accuracy constrains the time variation of the equation of state 
to below $w_a\lesssim0.5$ (the current best fit mirage value), crossing 
cannot be said to be tested significantly.  

Current distance data has 
insufficient leverage on the higher order kinematic quantities, such as 
the deceleration parameter $q(z)$.  It does definitely show, by this 
substantially nonparametric approach, that cosmic acceleration is occurring 
at low redshift.  The transition from acceleration to deceleration, however, 
could have happened at any redshift $z\gtrsim0.7$, or even 
not at all, according to this current data.  Future kinematic data extending 
to redshifts $z\gtrsim1$ are necessary to resolve all the issues of 
substantial time variation, phantom crossing, and the onset of 
acceleration.  Future applications of GP include projection of such 
constraints from future surveys, possibly allowing for spatial curvature, 
and tests of modified gravity, say, through growth vs.\ expansion.

\acknowledgments 

We thank Rollin Thomas for helpful discussions.  
This work has been supported by World Class University grant 
R32-2009-000-10130-0 through the National Research Foundation, Ministry 
of Education, Science and Technology of Korea and the Director, 
Office of Science, Office of High Energy Physics, of the U.S.\ Department 
of Energy under Contract No.\ DE-AC02-05CH11231.

\appendix 
\section{GP Mean Function \label{sec:appmean}} 

The GP formalism gives a clear formulation for derivation of the kinematic 
functions and their derivatives using the data, however there are some 
practical technical details that require care.  One of the important issues 
is the initial guess for the mean function.  The final results turn out 
not to be independent of this for an arbitrary choice, but can in fact 
retain some memory of the initial choice, biasing the results. 

One reason for this is because of the multiscale nature of the data for 
an arbitrary cosmology.
With two square-exponential-kernel hyperparameters, one for coherence 
length and one for the amplitude of deviations from the mean function, 
there exists limited freedom for the GP to track the deviations of the 
data from the input mean function redshift by redshift.  One solution is 
to use more hyperparameters for the GP covariance function
but this leads 
to a greater computational burden and the possibility of fitting the noise 
in the data rather than the cosmological signal. 

The residuals of the data around different mean functions can be quite 
different.  Certainly there is little expectation that any of these 
residuals are perfectly described by a Gaussian Process with a particular 
kernel.   It is therefore not unexpected for model predictions from 
different mean functions to be statistically inconsistent.

To give another view on this, consider the GP likelihood probability 
in more detail.  It contains 
three parts: the usual $\chi^2$, the determinant of the GP likelihood 
function penalizing overcomplexity, and a constant contribution involving 
the number of data points \cite{gpml},  
\begin{equation}
2 \ln p(y|f)= - y^T \Sigma_{00}(Z,Z)^{-1} y  - \ln\det\Sigma_{00}(Z,Z) 
- n \ln (2\pi)  \,,
\label{likelihood:eqn}
\end{equation} 
GP tries to find the best combination of first and second parts to get the 
highest likelihood.  That is, it tries to make $f$ as close as possible to 
the data $y$ by making minimum changes to the given mean function 
to get a reasonable $\chi^2$ and at the same time tries to keep the results 
smooth enough to get a high likelihood from the second term.  Note the 
second term is independent of the data and arises solely from the 
hyperparameters.  (This simplified explanation is somewhat complicated by 
the weighted integration over the hyperparameter space, but the basic 
flavor of it holds.)  

The ultimate model-independent input mean function is  the zero mean function. 
Here, however, we need a large $\sigma_f$ to bring $f$ close to $y$; to 
apply this ``correction'' to zero input over a large redshift range, without 
merely being a constant offset, requires a large coherence length $l$.  
While one might then succeed in fitting $f$ to $y$, this comes at the 
price of smoothing away the features and losing accurate reconstruction 
of $y'$ and $y''$.  Conversely, if we choose an input mean function that 
gives $f$ close to $y$ at several redshifts, then GP wants to keep 
$\sigma_f$ small to change the input function as little as possible.  
A small $\sigma_f$ effectively makes GP moot and so the reconstruction 
never strays far from the input function, with the results retaining 
memory of the input. 

We have verified these properties by investigating a large variety of 
mean functions: 1) zero mean function, 2) flat $\Lambda$CDM model with a 
fixed $\Omega_m$, 3) flat $\Lambda$CDM model with $\Omega_m$ as an added 
hyperparameter, 4) flat $w$CDM model with $\Omega_m$ and $w$ as added 
hyperparameters.  The issues raised above show up clearly. 

To get around these problems we want to use an input mean function with 
little cosmology dependence (for robust results), sensitivity to multiple 
scales (for flexibility in fitting $y$ well enough that the derivatives 
are reconstructed well), and few added hyperparameters (for computational 
tractability).  The solution we have adopted after extensive testing is 
an iterated smoothing approach, building on the method developed in 
\cite{Shafieloo06,Shafieloo07,Shafieloo10}. 

We emphasize that the smoothing is only to generate an initial mean 
function.  The data is smoothed over a scale $\Delta$ in $\ln(1+z)$, 
after ``pre-whitening'' using an initial guess $d_L^g$ \cite{Shafieloo10}: 
\begin{eqnarray} 
\label{eq:bg}
&\,&\ln d_L(z,\Delta)^{\rm s}-\ln d_L(z)^g = N(z)\times\\ 
&\quad & \sum_i \frac{\ln d_L(z_i)- \ln d_L(z_i)^g }{\sigma^2_{d_L(z_i)}} \ {\rm exp} \left [- \frac{\ln^2 \left
( \frac{1+z_i}{1+z} \right ) }{2 \Delta^2} \right ],  \nonumber \\
&\,&N(z)^{-1}=\sum_i {\rm exp} \left
[- \frac{\ln^2 \left ( \frac{1+z_i}{1+z} \right ) }{2 \Delta^2} \right ] \frac{1}{\sigma^2_{d_L(z_i)}} ~. 
\end{eqnarray}

This procedure is then iterated, with the output of one iteration serving 
as the initial guess of the successive iteration.  The final reconstructed 
results have been shown to be independent of the first initial guess 
\cite{Shafieloo06,Shafieloo07,Shafieloo10}.  Because the final iteration 
is a smooth function, we can take the derivatives of the mean functions 
as required for computing $f'$ and $f''$.  To incorporate many scales 
we actually use a set of 5 initial guess mean functions, iterating each 
one independently.  The scatter in the final mean functions or the 
appropriate derivatives then is added, weighted by likelihood, 
as an additional uncertainty, in the statistical sense of the mean squared 
error known as risk: the quadratic sum of the mean dispersion and the 
dispersion in the mean.
We find that stopping the iteration procedure 
when the $\chi^2$ of each lies within $\Delta\chi^2=2.3$ of each other 
(equivalent to $1\sigma$ for 2 degrees of freedom, hence as much dispersion 
as from $l$ and $\sigma_f^2$) gives robust results, as seen from the 
accurate reconstructions achieved for the \lcdm, mirage, and kink 
simulations.  In order to capture the true data it is important to have 
inputs that can cross the true cosmology over different redshift ranges, 
so the 5 initial guesses cover a wide range of behaviors, 
$\Lambda$CDM cosmologies with
$(\om,\Omega_\Lambda)=(1,0), (0,1), (0,0), (0.3,0.7), (0.5,0.5)$.  
Note that no additional hyperparameters are introduced. 

This iterated smoothing set approach to the mean function has the desired 
properties of not relying on a specific cosmological form and having 
freedom from memory of the initial guess, while allowing the GP formalism 
to balance the different terms in the likelihood and give accurate 
reconstructions with well characterized errors.  The tests run for 
reconstruction of the different cosmologies as shown in 
Figs.~\ref{fig:hrecon} and \ref{fig:qrecon} demonstrate its success.

\section{Hyperparameter Distribution} \label{sec:apphyper} 

The hyperparameters of the covariance function are another ingredient 
for the GP reconstruction.  Each set of hyperparameters, in our case 
$\sigma^2_f$ and $l$, gives rise to a particular likelihood by 
Eq.~(\ref{likelihood:eqn}), for each of the five mean functions. 

The posterior distribution is derived using Bayes' theorem
\begin{equation}
P(i,\sigma^2_f,l) = \frac{L(i,\sigma^2_f,l) p(i)p(\sigma^2_f)p(l) }{\sum_{i=1}^5  \int L(i,\sigma^2_f,l) p(i)p(\sigma^2_f)p(l) d \ln{\sigma^2_f} d\ln{l}},
\end{equation}
where $i$ is the index for the five mean functions, $p(i)=1/5$,
and the other priors are flat in the logarithm for 
$10^{-5} \leq \sigma^2_f \leq 1 $ and  $10^{-2} \leq l \leq 1.6$.  
The lower limit on $\sigma^2_f$ and upper limit of $l$ do impose
non-trivial truncation of the likelihood surface as discussed below. 
Note that setting a minimum $l=10^{-2}$ is equivalent to imposing a
blurring on the square-exponential kernel, and prevents fitting to 
merely noise in the data.  
Realizations of $f$, $f'$, and $f''$ are drawn from the Gaussian
Process models represented by this posterior.

\begin{figure}
  \begin{center}{
\includegraphics[angle=-90,width=\columnwidth]{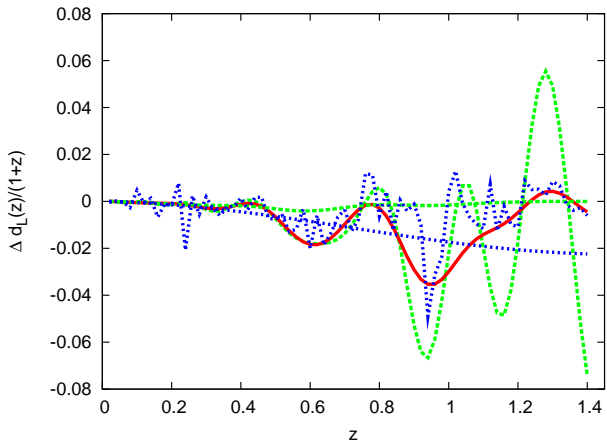}
  }
  \end{center}
  \caption{Schematic plot of the effects of $\sigf$ and $l$ on the 
reconstruction function $f(z)$.  The solid red curve has $l=0.1$ and 
$\sigf^2=0.001$ and will serve as a reference.  The two dark, blue dotted 
lines 
show the impact of changing $l$, with $l=0.01$ (wiggly line) and $l=1.0$ 
(nearly smooth line), keeping $\sigf$ unchanged.  The two light, green dashed
lines show the impact of changing $\sigf$, with $\sigf^2=0.1$ (increased 
amplitude of deviations in $f(z)$) and $\sigf^2=0.00001$ (decreased 
amplitude), keeping $l$ at the reference value.  For very small values of 
$\sigf^2$, GP makes very little contribution (near zero modification at all 
scales), while for high values of $l$ possible features of the data might 
be smoothed out. 
}
\label{fig:schematic}
\end{figure}

Figure~\ref{fig:schematic} illustrates the role of $\sigf^2$ and $l$ 
in the reconstruction.  Basically $\sigf$ acts to set the amplitude 
for deviations from the mean function and $l$ controls the wiggliness, 
or coherence scale.

As discussed in Appendix~\ref{sec:appmean}, a small value of $\sigma^2_f$ 
represents little contribution of GP to the reconstruction process, i.e.\ 
the result is basically just the mean function.  Large values of $l$ 
smooth over features in the data and basically give merely an offset that 
could be absorbed in the amplitude.  If we included arbitrarily small 
$\sigf^2$ or large $l$ in the hyperparameter ranges, these regions of the 
space would give nearly identical likelihoods and dilute the overall 
probabilities, biasing the results toward the mean function.  To avoid 
this situation we impose the lower limit $\log\sigf^2 \geq -5$ (i.e.\ ignoring 
models changing the mean function by less than 0.7\%) and the upper limit 
$l \leq 1.6$ (i.e.\ the range of the data).  
We have checked that the final best fits for the hyperparameters are not 
significantly affected by small variations in the priors.


\end{document}